\documentclass[twocolumn,pra,aps,superscriptaddress,floatfix]{revtex4}
\usepackage{amssymb}
\usepackage{amsfonts}
\usepackage{amsmath}
\usepackage{graphicx}

\begin{document}

\title{Perturbation theory for operational quantum non-Markovianity}
\author{Mariano Bonifacio}
\affiliation{Instituto Balseiro, Universidad Nacional de Cuyo, Avenida E. Bustillo Km
9.5, (8400) Bariloche, Argentina}
\author{Adri\'{a}n A. Budini}
\affiliation{Consejo Nacional de Investigaciones Cient\'{\i}ficas y T\'{e}cnicas
(CONICET), Centro At\'{o}mico Bariloche, Avenida E. Bustillo Km 9.5, (8400)
Bariloche, Argentina, and Universidad Tecnol\'{o}gica Nacional (UTN-FRBA),
Fanny Newbery 111, (8400) Bariloche, Argentina}

\begin{abstract}
The definition of memory in operational approaches to quantum
non-Markovianity depends on the statistical properties of different sets of
outcomes related to successive measurement processes performed over the
system of interest. Using projectors techniques we develop a perturbation
theory that enables to expressing both joint probabilities and outcome
correlations in terms of the unperturbed system density matrix propagator.
This object defines the open system dynamics in absence of measurement
processes. Successive series terms, which are scaled by the
system-environment interaction strength, consist in a convolution structure
involving system propagators weighted by higher order bath correlations. The
formalism is corroborated by studying different dynamics that admit an exact
description. Using the perturbative approach, unusual memory effects induced
by the interplay between the system-environment interaction and measurement\
processes are found in finite temperature reservoirs.
\end{abstract}

\date{\today }
\maketitle

\section{Introduction}

Most features characterizing an open quantum system dynamics can be
recovered from a \textit{perturbative} approach to the full
system-environment dynamics. For example, the Born-Markov approximation
(BMA) \cite{cohenbook,breuerbook,mandel} is able to describe paradigmatic
phenomena like decoherence and dissipation. Even more, quantum memory
effects were originally related to departures from this \textquotedblleft
first-order\textquotedblright\ approximation \cite{vega,wiseman}. This
association relies on the local-in-time property of the system evolution.

Over the past decade, the previous point of view was reviewed significantly.
Instead of the BMA, the hallmark of quantum Markovianity becomes the theory
of quantum dynamical semigroups \cite{alicki}. In this alternative scenario 
\cite{BreuerReview,plenioReview} memory effects are determined from the
system density matrix \textit{propagator}, which in fact encode different
departures that the system dynamics may develop with respect to a
\textquotedblleft Markovian\textquotedblright\ Lindblad dynamics~\cite%
{BreuerFirst,cirac,rivas,breuerDecayTLS,fisher,fidelity,dario,mutual,geometrical, DarioSabrina,brasil,sabrina,canonicalCresser,cresser,paris,Acin,indu}%
.

Alternative\textit{\ operational }approaches \cite{ModiOperational}\ to
quantum non-Markovianity have been introduced recently \cite%
{modi,pollock,pollockInfluence,budiniCPF,budiniChina,budiniBrasil}. Instead
of focusing on mappings from density operators to density operators, the
presence of memory is determined from the statistical properties of
different outcomes obtained from successive measurement processes performed
during the system evolution. Consistence with the classical definition of
non-Markovianity is achieved. In fact, given a sequence of measurement
outcomes, Markovianity can be checked through the corresponding conditional
probabilities \cite{modi}. In addition and in contrast to previous
non-operational approaches, any possible dynamical departure from BMA
renders the dynamic non-Markovian~\cite{budiniCPF}. Experimental setups for
measuring memory in an operational way were implemented recently in Refs.~%
\cite{budiniChina,budiniBrasil}.

The definition of quantum non-Markovianity from an operational perspective
leads to an intrinsic dependence of memory effects on measurement processes,
which leads to a richer structure when compared to the classical
(incoherent) case \cite{pollock,pollockInfluence,budiniChina}. On the other
hand, in contrast to non-operational approaches where memory effects can be
related to the system propagator, it is not known which physical object (or
structure) may play the same role in these operational approaches \cite%
{budiniBrasil,goan}. In fact, the system dynamic between successive
measurements, due to non-Markovian effects, cannot in general be described
through a unique system propagator. This property has its physical origin in
the modification or dependence of the bath state on system outcomes~\cite%
{budiniCPF}. The main goal of this paper is to provide a rigorous answer to
this problem.

We characterize the structure that determines memory effects in operational
approaches to quantum non-Markovianity. The study is valid for arbitrary
system-environment interactions and relies on a perturbation theory
formulated with projector operator techniques \cite{breuerbook,projectors}.
The formalism is developed in the case where three system measurement
processes are performed, being applied to both joint probabilities \cite%
{modi} and a conditional past-future (CPF) correlation \cite{budiniCPF}. In
order to understand the intrinsic differences between operational and
non-operational approaches, both kind of statistical objects are written as
a function of the\textit{\ unperturbed system propagator}, which defines the
open system dynamics in absence of measurement processes. The projector
approach naturally leads to an expansion series in the system-environment
coupling strength. We found that successive order contributions consist in a
convolution term involving two system propagators weighed by higher order
bath correlations. This structure arises for both quantum and classical
environmental fluctuations. These finding generalize the results found in 
\cite{budiniBrasil}, which were derived for specific system-bath interaction
Hamiltonians. The validity of the formalism is confirmed by studying
different dynamics that admit an exact treatment such as dephasing and decay
in a bosonic bath at zero temperature. As an application, we study memory
effects in the thermalization of a two-level system, showing that unusual
memory effects may be induced by raising the environment temperature.
Generalization to arbitrary number of measurement processes follows
straightforwardly from the present results.

The paper is outlined as follows. In Sec. II we review how memory effects
can be determined from joint probabilities and the CPF correlation. In Sec.
III we develop the perturbation theory for both classical and quantum
environment fluctuations. In Sec. IV we apply the perturbation theory to
dynamics that admit an exact treatment. In addition, we study memory effects
induced by thermal reservoirs. In Sec. V we provide the Conclusions.
Auxiliary calculation details are presented in the Appendix.

\section{Operational memory witnesses}

Memory effects in open quantum systems can be determined by subjecting the
system to successive measurement processes and checking if the corresponding
probability structure satisfies the usual Markovian definition \cite{modi}.
It is simple to realize that a minimal number of three system observations
is necessary to detect memory effects. Denoting with $x\rightarrow
y\rightarrow z$ the successive measurement outcomes, their joint probability 
$P(z,y,x)$\ can be written as%
\begin{equation}
P(z,y,x)=P(z|y,x)P(y|x)P(x),  \label{3Joint}
\end{equation}%
where in general, $P(b|a)$ denotes the conditional probability of $b$ given $%
a.$ Markovianity is defined by the equality $P(z,y,x)\overset{M}{=}%
P(z|y)P(y|x)P(x),$ that is, $P(z|y,x)\overset{M}{=}P(z|y).$ This property
can easily be rewritten in terms of a \textit{conditional} past-future
independence, leading to the condition $P(z,x|y)\overset{M}{=}P(z|y)P(x|y).$
This last formulation can be checked with a CPF correlation \cite{budiniCPF},%
\begin{equation}
C_{pf}(t,\tau )|_{y}=\sum_{zx}[P(z,x|y)-P(z|y)P(x|y)]O_{z}O_{x}.
\label{CPFDefinition}
\end{equation}%
Thus, Markovianity implies $C_{pf}(t,\tau )|_{y}=0,$ while $C_{pf}(t,\tau
)|_{y}\neq 0$ witnesses memory effects. In this equation, the sequence $%
\{x\}\rightarrow y\rightarrow \{z\}$ defines the outcomes at each stage,
while $\{O_{z}\}$ and $\{O_{x}\}$ are the corresponding observables at the
initial and final (past and future) observation times. The outcome $y$ gives
the conditional character of the correlation. The parameters $t$ and $\tau $
denote the time intervals between the first and second, and between the
second and third measurements, respectively.

Both operational memory witnesses [Eqs. (\ref{3Joint}) and (\ref%
{CPFDefinition})] can be mapped between them. Using that $%
P(z,x|y)=P(z,y,x)/P(y),$ $P(z|y)=P(z,y)/P(y)$ and $P(x|y)=P(y,x)/P(y),$ we
get the equivalent expression%
\begin{equation}
C_{pf}(t,\tau )|_{y}=\sum_{zx}\frac{O_{z}O_{x}}{P^{2}(y)}%
[P(z,y,x)P(y)-P(z,y)P(y,x)].  \label{CPF3Joint}
\end{equation}
Here, all statistical objects can be written in terms of the joint
probability $P(z,y,x).$ In fact, $P(z,y)=\sum_{x}P(z,y,x),$ $%
P(y,x)=\sum_{z}P(z,y,x)$ and $P(y)=\sum_{z,x}P(z,y,x).$

In a quantum regime, joint probabilities as well as the CPF correlation
intrinsically depend on the chosen observables. Here they are defined
through a set of measurement operators denoted as $\{\Omega _{x}\},$ $%
\{\Omega _{y}\},$ and $\{\Omega _{z}\},$ being normalized to the system
identity matrix, $\sum_{z}\Omega _{z}^{\dagger }\Omega _{z}=\sum_{y}\Omega
_{y}^{\dagger }\Omega _{y}=\sum_{x}\Omega _{x}^{\dagger }\Omega _{x}=\mathrm{%
I}.$ For simplicity, the intermediate measurement is assumed a projective
one \cite{modi,budiniCPF}, that is, $\Omega _{y}^{\dagger }\Omega
_{y}=\Omega _{y}.$

For the explicit calculation of $P(z,y,x)$ or $C_{pf}(t,\tau )|_{y}$\ we
must define the evolution of the system-environment arrange between
measurements. Both (total) unitary dynamics and stochastic Liouville
dynamics are considered.

\subsection{Unitary system-environment dynamics}

First, we assume that the system and the environment are described by a
unitary evolution with Hamiltonian $H_{T}.$ The total density matrix $\rho
_{t}^{se}$ evolves as%
\begin{equation}
\frac{d}{dt}\rho _{t}^{se}=\mathcal{L}_{se}(t)[\rho _{t}^{se}],\ \ \ \ \ \ \ 
\mathcal{L}_{se}(t)[\bullet ]=-i[H_{T}(t),\bullet ].
\end{equation}%
As usual, the total Hamiltonian is written as $H_{T}(t)=H_{s}+H_{e}+H_{I}.$
Each contribution corresponds to the system, the environment, and their
interaction Hamiltonian, respectively. The previous equation can be
integrated as $\rho _{t}^{se}=\mathcal{E}_{t,0}[\rho _{0}^{se}],$ where the
bipartite propagator is%
\begin{equation}
\mathcal{E}_{t_{b},t_{a}}\equiv \left\lceil \exp
\int_{t_{a}}^{t_{b}}dt^{\prime }\mathcal{L}_{se}(t^{\prime })\right\rceil .
\end{equation}%
Here $\left\lceil \cdots \right\rceil $ denotes a time ordering operation,
which is necessary due to the dependence of $\mathcal{L}_{se}(t)$ on time.
This case arises, for example, when working in an interaction representation
or when the system is submitted to an external time dependent field.

The system density matrix follows from a partial trace over the
environmental degrees of freedom, $\rho _{t}=\mathrm{Tr}_{e}(\rho
_{t}^{se}). $ Thus,%
\begin{equation}
\rho _{t}=\Lambda _{t,t_{0}}[\rho _{0}]\equiv \mathrm{Tr}_{e}(\mathcal{E}%
_{t,t_{0}}[\rho _{0}\otimes \sigma _{e}]),  \label{PropaCuantico}
\end{equation}%
where $\Lambda _{t,t_{0}}$ is the system density matrix propagator. For
simplicity, we assume $t_{0}=0$ and\textit{\ separable initial conditions}, $%
\rho _{0}^{se}=\rho _{0}\otimes \sigma _{e}.$

From standard quantum measurement theory, the expression for the 3-joint
probability is \cite{suple}%
\begin{equation}
P(z,y,x)=\mathrm{Tr}_{se}(E_{z}\mathcal{E}_{t+\tau ,t}[\rho _{y}\otimes 
\mathrm{Tr}_{s}(E_{y}\mathcal{E}_{t,0}[\tilde{\rho}_{x}\otimes \sigma
_{e}])]),  \label{P3Quantum}
\end{equation}%
where $\tilde{\rho}_{x}\equiv \Omega _{x}\rho _{0}\Omega _{x}^{\dagger }$
and $E_{i}\equiv \Omega _{i}^{\dag }\Omega _{i}.$ Furthermore, $\rho
_{y}=E_{y}$ is the (collapsed) system state after the second measurement. We
notice that in Eq.~(\ref{P3Quantum}) the evolution in the interval $(0,t)$
can be written in terms of the unperturbed system propagator $\Lambda _{t,0}$
defined in Eq.~(\ref{PropaCuantico}). Nevertheless, this object is
insufficient to describe the dynamics in the interval $(t,t+\tau )$ because
the initial bath state does not remain unchanged, $\sigma _{e}\rightarrow 
\mathrm{Tr}_{s}(E_{y}\mathcal{E}_{t,0}[\tilde{\rho}_{x}\otimes \sigma
_{e}]). $ In fact, this feature is a witness of memory effects \cite%
{budiniCPF}\ whose description, for arbitrary system-environment
interactions, is performed in the following section.

From Eq. (\ref{P3Quantum}), the CPF correlation [Eq. (\ref{CPF3Joint})] can
be written as%
\begin{widetext}%
\begin{equation}
C_{pf}(t,\tau )|_{y}=\frac{1}{P^{2}(y)}\sum_{zx}O_{z}O_{x}\mathrm{Tr}%
_{se}(E_{z}\mathcal{E}_{t+\tau ,t}[\rho _{y}\otimes \mathrm{Tr}_{s}(E_{y}%
\mathcal{E}_{t,0}[\tilde{\rho}_{yx}\otimes \sigma _{e}])]),
\label{CPFUnitary}
\end{equation}%
\end{widetext}%
where the\ auxiliary system matrix $\tilde{\rho}_{yx}$ is defined as $\tilde{%
\rho}_{yx}\equiv \tilde{\rho}_{x}\ P(y)-\tilde{\rho}\ P(y,x),$\ being $%
\tilde{\rho}\equiv \sum_{x^{\prime }}\tilde{\rho}_{x^{\prime }}.$
Explicitly, it reads%
\begin{equation}
\tilde{\rho}_{yx}=\tilde{\rho}_{x}\ \mathrm{Tr}_{s}(E_{y}\Lambda _{t,0}[%
\tilde{\rho}])-\tilde{\rho}\ \mathrm{Tr}_{s}(E_{y}\Lambda _{t,0}[\tilde{\rho}%
_{x}]),  \label{RhoYX}
\end{equation}%
Similarly, the probability $P(y)$ is given by%
\begin{equation}
P(y)=\sum_{x^{\prime }}\mathrm{Tr}_{s}(E_{y}\Lambda _{t,0}[\tilde{\rho}%
_{x^{\prime }}]).  \label{P(y)}
\end{equation}%
From the previous two expressions, we notice that both $\tilde{\rho}_{yx}$
and $P(y)$ can be written in terms of the unperturbed system propagator $%
\Lambda _{t,0}.$ On the other hand, it is simple to show that the matrix $%
\tilde{\rho}_{yx}$ never vanishes. In fact, after a simple algebra the
condition $\tilde{\rho}_{yx}=0$ leads to the incongruence $\tilde{\rho}%
_{x}/P(x)=\tilde{\rho}.$

We notice that Eq.~(\ref{CPFUnitary}), disregarding the sum operation and
under the replacement $\tilde{\rho}_{yx}\rightarrow \tilde{\rho}_{x},$ has
the same structure as Eq.~(\ref{P3Quantum}). This similitude allows us to
formulate a perturbation theory that straightforwardly applies to both kinds
of objects. The same relation is also valid for higher statistical objects
(Sec. IV-C).

\subsection{Stochastic Liouville dynamics}

In addition, we deal with the case in which the open system evolution is
defined by a stochastic Liouville dynamics,%
\begin{equation}
\frac{d}{dt}\rho _{t}^{st}=\mathcal{L}_{st}(t)[\rho _{t}^{st}],\ \ \ \ \ 
\mathcal{L}_{st}(t)[\bullet ]=-i[H_{st}(t),\bullet ].  \label{EleStochastic}
\end{equation}%
This equation can be integrated as $\rho _{t}^{st}=\mathcal{E}%
_{t,0}^{st}\rho _{0},$ where the stochastic propagator is%
\begin{equation}
\mathcal{E}_{t_{b},t_{a}}^{st}\equiv \left\lceil \exp
\int_{t_{a}}^{t_{b}}dt^{\prime }\mathcal{L}_{st}(t^{\prime })\right\rceil .
\end{equation}%
As before, $\left\lceil \cdots \right\rceil $ denotes a time ordering
operation. The system density matrix $\rho _{t}=\overline{\rho _{t}^{st}}$
follows after averaging over realizations (over bar symbol) of the
stochastic Liouville superoperator $\mathcal{L}_{st}(t).$ Thus, 
\begin{equation}
\rho _{t}=\Lambda _{t,t_{0}}[\rho _{0}]\equiv \overline{\mathcal{E}%
_{t,t_{0}}^{st}}[\rho _{0}],  \label{PropaClasico}
\end{equation}%
where for simplicity we assumed that the initial system state $\rho _{0}$\
is \textit{uncorrelated} from the noise fluctuations. As before, $\Lambda
_{t,t_{0}}$ $(t_{0}=0)$ is the system density matrix propagator.

From quantum measurement theory, it is possible to obtain \cite{suple} 
\begin{equation}
P(z,y,x)=\overline{\mathrm{Tr}_{s}(E_{z}\mathcal{E}_{t+\tau ,t}^{st}[\rho
_{y}])\mathrm{Tr}_{s}(E_{y}\mathcal{E}_{t,0}^{st}[\tilde{\rho}_{x}])},
\label{P3Noise}
\end{equation}%
where, as before, $\tilde{\rho}_{x}=\Omega _{x}\rho _{0}\Omega _{x}^{\dagger
}$ and $E_{i}=\Omega _{i}^{\dag }\Omega _{i}.$ Similarly to the unitary
case, here the (average) evolution in the interval $(0,t)$ can be written in
terms of the unperturbed propagator (\ref{PropaCuantico}), but it is unable
to describe the dynamics in the interval $(t,t+\tau )$ because the system
state at time $t$ is a random one, being correlated with the environmental
fluctuations.

From Eq. (\ref{P3Noise}) it is possible to write the CPF correlation [Eq. (%
\ref{CPF3Joint})] as 
\begin{equation}
C_{pf}(t,\tau )|_{y}=\!\sum_{zx}\frac{O_{z}O_{x}}{P^{2}(y)}\overline{\mathrm{%
Tr}_{s}(E_{z}\mathcal{E}_{t+\tau ,t}^{st}[\rho _{y}])\mathrm{Tr}_{s}(E_{y}%
\mathcal{E}_{t,0}^{st}[\tilde{\rho}_{yx}])},  \label{CPFNoise}
\end{equation}
where $\tilde{\rho}_{yx}$\ and $P(y)$ can be read from Eqs. (\ref{RhoYX})
and (\ref{P(y)}), respectively, with the propagator $\Lambda _{t,0}$ defined
by Eq.~(\ref{PropaClasico}).

Here, we can see that Eqs. (\ref{P3Noise}) and (\ref{CPFNoise}) present a
similar structure, and both expressions can be related under the same
mapping that connects Eqs. (\ref{P3Quantum}) and (\ref{CPFUnitary}).

\section{Perturbation theory}

In non-operational memory approaches, memory effects are mainly determined
from the unperturbed system density propagator. Thus, we develop a
perturbation theory where this object remains as an input of the formalism.
For both unitary system-environment interactions as well as stochastic
Liouville dynamics the formalism is developed using projector techniques,
which allow us to find exact series expansions of both the joint probability 
$P(z,y,x)$ and the CPF correlation $C_{pf}(t,\tau )|_{y}.$

\subsection{Unitary system-environment dynamics}

We introduce standard projectors \cite{projectors}%
\begin{equation}
\mathcal{P}[\rho _{se}]=\mathrm{Tr}_{e}(\rho _{se})\otimes \sigma _{e},\ \ \
\ \ \mathcal{Q}[\rho _{se}]=\rho _{se}-\mathrm{Tr}_{e}(\rho _{se})\otimes
\sigma _{e},  \label{QProjectores}
\end{equation}%
where $\sigma _{e}$\ is a reference state of the bath. They satisfy $%
\mathcal{P}+\mathcal{Q}=\mathrm{I}_{se},$ where $\mathrm{I}_{se}$ is the
bipartite identity matrix. Introducing the operator $\mathrm{I}_{se}$ in
front of each propagator $\mathcal{E}$ in Eq.~(\ref{P3Quantum}), it follows%
\begin{equation}
\begin{array}[b]{r}
P(z,y,x)\!=\!\mathrm{Tr}_{se}(E_{z}\mathcal{PE}_{t+\tau ,t}[\rho
_{y}\!\otimes \!\mathrm{Tr}_{s}(E_{y}\mathcal{PE}_{t,0}[\tilde{\rho}%
_{x}\otimes \sigma _{e}])]) \\ 
\\ 
\!+\mathrm{Tr}_{se}(E_{z}\mathcal{PE}_{t+\tau ,t}[\rho _{y}\!\otimes \!%
\mathrm{Tr}_{s}(E_{y}\mathcal{QE}_{t,0}[\tilde{\rho}_{x}\otimes \sigma
_{e}])]).%
\end{array}
\label{Sol}
\end{equation}%
In deriving this expression we used that $\mathrm{Tr}_{e}(\mathcal{Q}[\rho
_{se}])=0,$ equality valid for arbitrary system-environment state $\rho
_{se}.$ We notice that the first line in the previous equation can be
written in terms of two system propagators [Eq. (\ref{PropaCuantico})]
between two arbitrary times, $\Lambda _{t,t^{\prime }}[\rho ]=\mathrm{Tr}%
_{e}(\mathcal{E}_{t,t^{\prime }}[\rho \otimes \sigma _{e}]).$ In the second
line, as usual, we note that for separable initial conditions the irrelevant
part\ in the projector technique can be integrated as (see Appendix)%
\begin{equation}
\mathcal{QE}_{t_{b},t_{a}}=\int_{t_{a}}^{t_{b}}dt^{\prime }\mathcal{G}%
_{t,t^{\prime }}\mathcal{QL}_{se}(t^{\prime })\mathcal{PE}_{t^{\prime
},t_{a}},  \label{IrrelevantPart}
\end{equation}%
where%
\begin{equation}
\mathcal{G}_{t,t^{\prime }}\equiv \left\lceil \exp \int_{t^{\prime
}}^{t}d\tau ^{\prime }\mathcal{QL}_{se}(\tau ^{\prime })\right\rceil .
\end{equation}%
Therefore, the contribution proportional to $\mathcal{QE}_{t_{b},t_{a}}^{st}$
in Eq.~(\ref{Sol}) can also be written in terms of the unperturbed
propagator $\mathcal{PE}_{t^{\prime },t_{a}}[\rho \otimes \sigma _{e}]$
[Eq.~(\ref{PropaCuantico})]. On the other hand, the relevant part (in the
second line) can also be integrated as (see Appendix)%
\begin{equation}
\mathcal{PE}_{t_{b},t_{a}}=\mathcal{PE}_{t_{b},t_{a}}\mathcal{P}%
\!+\!\int_{t_{a}}^{t_{b}}dt^{\prime }\mathcal{PE}_{t_{b},t^{\prime }}%
\mathcal{P\mathcal{L}}_{se}(t^{\prime })\mathcal{G}_{t^{\prime },t_{a}}%
\mathcal{Q}.  \label{RelevantPart}
\end{equation}%
This expression is of central importance for the developing of the formalism
because it enables to characterize the projected system dynamics in terms of
the unperturbed propagator even when considering \textit{arbitrary initial
environment states}.

Introducing explicitly Eqs. (\ref{IrrelevantPart}) and (\ref{RelevantPart})
in $P(z,y,x),$ using that $\mathcal{PQ}=0$ and after some algebra, from Eq. (%
\ref{Sol}) we get%
\begin{widetext}%
\begin{equation}
P(z,y,x)\!=\!\mathrm{Tr}_{s}(E_{z}\Lambda _{t+\tau ,t}[\rho _{y}])\mathrm{Tr}%
_{s}(E_{y}\Lambda _{t,0}[\tilde{\rho}_{x}])+\int_{0}^{\tau }\!\!\!d\tau
^{\prime }\int_{0}^{t}\!\!\!dt^{\prime }\mathrm{Tr}_{s}(E_{z}\Lambda
_{t+\tau ,t+\tau ^{\prime }}\mathrm{Tr}_{e}(\tilde{\Phi}_{t+\tau ^{\prime
},t}^{se}[\rho _{y}\otimes \mathrm{Tr}_{s}(E_{y}\Phi _{t,t^{\prime
}}^{se}\Lambda _{t^{\prime },0}[\tilde{\rho}_{x}]\otimes \sigma _{e})])),
\label{3JointQuantum}
\end{equation}%
\end{widetext}%
where for shortening the expression we introduced the system-environment
superoperators%
\begin{equation}
\tilde{\Phi}_{t_{b},t_{a}}^{se}=\mathcal{\mathcal{L}}_{se}(t_{b})\mathcal{G}%
_{t_{b},t_{a}},\ \ \ \ \ \ \ \ \Phi _{t_{b},t_{a}}^{se}\equiv \mathcal{G}%
_{t_{b},t_{a}}\mathcal{QL}_{se}(t_{a}).  \label{BipartiteSUP}
\end{equation}%
The system propagator $\Lambda _{t,t^{\prime }}$ is defined by Eq. (\ref%
{PropaCuantico}).

Eq. (\ref{3JointQuantum}) is the main result of this section. It expresses $%
P(z,y,x)$ as a function of the unperturbed system propagator. We notice that
the first contribution corresponds to a Markovian limit, where $P(z,y,x)%
\overset{M}{=}P(z|y)P(y|x)P(x)$ with $P(z|y)\overset{M}{=}\mathrm{Tr}%
_{s}(E_{z}\Lambda _{t+\tau ,t}[\rho _{y}])$ and $P(y|x)P(x)\overset{M}{=}%
\mathrm{Tr}_{s}(E_{y}\Lambda _{t,0}[\tilde{\rho}_{x}])$ with $P(x)=\mathrm{Tr%
}_{s}(\tilde{\rho}_{x})=\mathrm{Tr}_{s}(E_{x}\rho _{0}).$ Consistently, the
second (integral) contribution takes into account memory effects, which in
turn answers our main motivation. It consists in a convolution structure
involving two unperturbed system propagators weighted by the
\textquotedblleft correlation\textquotedblright\ between the bipartite
operators $\tilde{\Phi}_{t_{b},t_{a}}^{se}$ and $\Phi _{t_{b},t_{a}}^{se}$
[Eq.~(\ref{BipartiteSUP})]. These objects can be written as a series in the
interaction strength [proportional to $\mathcal{\mathcal{L}}_{se}(t)],$
which follows from the expansion $\mathcal{G}_{t,t^{\prime }}=\mathrm{I}%
_{se}+\int_{t^{\prime }}^{t}d\tau _{1}\mathcal{QL}_{se}(\tau
_{1})+\int_{t^{\prime }}^{t}d\tau _{2}\int_{t^{\prime }}^{\tau _{2}}d\tau
_{1}\mathcal{QL}_{se}(\tau _{2})\mathcal{QL}_{se}(\tau _{1})+\cdots .$ Thus,
the non-Markovian contribution in Eq. (\ref{3JointQuantum}) can be written
as a series in the interaction strength, each term involving two system
propagators and high order bath correlations.

In order to lighten the structure of the perturbation series, we write it
for the CPF correlation. Using the similarity between Eqs.~(\ref{P3Quantum})
and (\ref{CPFUnitary}), it follows that the first (Markovian) term in Eq.~(%
\ref{3JointQuantum}) does not contribute to $C_{pf}(t,\tau )|_{y}.$ In fact $%
\sum_{x}O_{x}\mathrm{Tr}_{s}(E_{y}\Lambda _{t,0}[\tilde{\rho}_{yx}])=0.$
Thus,\ consistently the CPF correlation only depends on the second integral
contribution, which in fact measures the memory effects. We get%
\begin{equation}
C_{pf}(t,\tau )|_{y}=\sum_{z,x}\frac{O_{z}O_{x}}{P^{2}(y)}%
\int_{0}^{t}dt^{\prime }\int_{0}^{\tau }d\tau ^{\prime }\Xi (z,x|y)[\tilde{%
\rho}_{yx}],  \label{Final}
\end{equation}%
where%
\begin{equation}
\Xi (z,x|y)[\bullet ]\equiv \mathrm{Tr}_{se}(E_{z}\tilde{\Upsilon}_{t+\tau
^{\prime },t}^{se}[\ \rho _{y}\otimes \mathrm{Tr}_{s}(E_{y}\Upsilon
_{t,t^{\prime }}^{se}[\bullet ])\ ]).  \label{Sigma}
\end{equation}%
This term defines the integrand in Eq. (\ref{3JointQuantum}). For notational
convenience here we introduced the superoperator%
\begin{equation}
\tilde{\Upsilon}_{t+\tau ^{\prime },t}^{se}[\bullet ]\equiv \Lambda _{t+\tau
,t+\tau ^{\prime }}\tilde{\Phi}_{t+\tau ^{\prime },t}^{se}[\bullet ],
\end{equation}%
and similarly%
\begin{equation}
\Upsilon _{t,t^{\prime }}^{se}[\bullet ]\equiv \Phi _{t,t^{\prime
}}^{se}[\Lambda _{t^{\prime },0}[\bullet ]\otimes \sigma _{e}].
\end{equation}

The final expression (\ref{Final}) enables to perform a perturbative theory
for the CPF correlation developed as a series in terms of the
system-environment interaction strength. In fact, expansion of the
superoperators $\tilde{\Phi}_{t_{b},t_{a}}^{se}$ and $\Phi
_{t_{b},t_{a}}^{se}$ [Eq. (\ref{BipartiteSUP})] in powers of $\mathcal{%
\mathcal{L}}_{se}(t)$ leads to%
\begin{equation}
\Xi (z,x|y)[\bullet ]=\sum_{n=1}^{\infty }\Xi ^{(n)}(z,x|y)[\bullet ],
\label{SerieSigma}
\end{equation}%
where the index $n$ labels the bath correlation order that appears in each
term. For example, to first order $\tilde{\Phi}_{t_{b},t_{a}}^{se}=\mathcal{%
\mathcal{L}}_{se}(t_{b})\mathcal{G}_{t_{b},t_{a}}\simeq \mathcal{\mathcal{L}}%
_{se}(t_{b}),$ and $\Phi _{t_{b},t_{a}}^{st}=\mathcal{G}_{t_{a},t_{b}}^{st}%
\mathcal{QL}_{se}(t_{a})\simeq \mathcal{QL}_{se}(t_{a})=\mathcal{L}%
_{se}(t_{a}),$ where the last equality relies on the usual assumption $%
\mathcal{PL}_{se}(t_{a})\mathcal{P}=0.$ The first not null order is weighted
by the bath correlations $\mathrm{Tr}_{e}(\mathcal{L}_{se}(t)\mathcal{L}%
_{se}(t^{\prime })\sigma _{e}),$ which in turn weights the integral between
both system propagators $\Lambda _{t+\tau ,t+\tau ^{\prime }}$ and $\Lambda
_{t^{\prime },0}.$ This structure is similar to that found in Ref. \cite%
{budiniBrasil} for models that admit an exact analytic calculation.

In order to explicitly visualize the previous structure, we consider the
bipartite Hamiltonian%
\begin{equation}
H_{T}(t)=\sum_{\mu }S_{t}^{\mu }\otimes B_{t}^{\mu }.
\end{equation}%
Assuming, as usual, that expectation values of the bath operators are null, $%
\mathrm{Tr}_{e}(B_{t}^{\mu }\sigma _{e})=0,$ and considering Hermitian
operators, from Eq. (\ref{Sigma}) we get%
\begin{equation}
\begin{array}{r}
\Xi ^{(1)}(z,x|y)[\bullet ]=\sum\limits_{\mu ,\nu }\Big{\{}\mathrm{Tr}_{s}%
\Big{(}E_{z}\Lambda _{t+\tau ,t+\tau ^{\prime }}[\rho _{y}S_{t+\tau ^{\prime
}}^{\mu }]\Big{)}-c.c.\Big{\}} \\ 
\\ 
\!\!\!\!\!\!\!\!\times \Big{\{}\chi _{\mu \nu }(\tau ^{\prime }+t^{\prime
})\ \mathrm{Tr}_{s}\Big{(}E_{y}S_{t-t^{\prime }}^{\nu }\Lambda _{t^{\prime
},0}[\bullet ]\Big{)}-c.c.\Big{\}},%
\end{array}%
\end{equation}%
where the bath correlations are defined as $\chi _{\mu \nu }(t,t^{\prime
})\equiv \mathrm{Tr}_{e}(B_{t}^{\mu }B_{t^{\prime }}^{\nu }\sigma _{e}).$
For simplicity they are assumed stationary, $\chi _{\mu \nu }(t+\tau
^{\prime },t-t^{\prime })=\chi _{\mu \nu }(\tau ^{\prime }+t^{\prime }).$
Higher order terms include higher bath correlations that involve a higher
number of bath operators. For bosonic environments, $\Xi
^{(n)}(z,x|y)[\bullet ]$ involves a product of $n$ correlations $\chi _{\mu
\nu }(t,t^{\prime }).$ When $\chi _{\mu \nu }(\tau ^{\prime }+t^{\prime
})\approx \delta (\tau ^{\prime }+t^{\prime }),$ the double time integral $%
\int_{0}^{t}dt^{\prime }\int_{0}^{\tau }d\tau ^{\prime }$ of the successive
series terms vanishes, recovering consistently a Markovian limit.

We remark that the exact expressions (\ref{3JointQuantum}) and (\ref{Final})
explicitly depend on the unperturbed propagator $\Lambda _{t_{b},t_{a}}.$
This object, when is not available in an exact analytical way, using
standard tools \cite{breuerbook,projectors}, can be approximated to the same
order as the joint probability or CPF correlation.

\subsection{Stochastic Liouville dynamics}

The previous perturbation theory can also be developed for the case of
stochastic Liouville dynamics, Eqs.~(\ref{P3Noise}) and (\ref{CPFNoise}).
Instead of the projectors (\ref{QProjectores}), here they are defined as%
\begin{equation}
\mathcal{P}[f_{st}]=\overline{f_{st}},\ \ \ \ \ \ \ \ \ \mathcal{Q}%
[f_{st}]=f_{st}-\overline{f_{st}},
\end{equation}%
where $f_{st}$ is an arbitrary functional of the noise fluctuations. After
introducing the identity $\mathcal{P}+\mathcal{Q}=1$ in Eq.~(\ref{P3Noise}),
we get%
\begin{equation}
\begin{array}{r}
P(z,y,x)=\overline{\mathrm{Tr}_{s}(E_{z}\mathcal{PE}_{t+\tau ,t}^{st}[\rho
_{y}])\mathrm{Tr}_{s}(E_{y}\mathcal{PE}_{t,0}^{st}[\tilde{\rho}_{x}])} \\ 
\\ 
+\overline{\mathrm{Tr}_{s}(E_{z}\mathcal{PE}_{t+\tau ,t}^{st}[\rho _{y}])%
\mathrm{Tr}_{s}(E_{y}\mathcal{QE}_{t,0}^{st}[\tilde{\rho}_{x}])}.%
\end{array}%
\end{equation}%
Using similar transformations and solutions as in the previous section, for
the joint probability we obtain%
\begin{widetext}%
\begin{equation}
P(z,y,x)\!=\!\mathrm{Tr}_{s}(E_{z}\Lambda _{t+\tau ,t}[\rho _{y}])\mathrm{Tr}%
_{s}(E_{y}\Lambda _{t,0}[\tilde{\rho}_{x}])+\int_{0}^{\tau }d\tau ^{\prime
}\int_{0}^{t}dt^{\prime }\overline{\mathrm{Tr}_{s}(E_{z}\Lambda _{t+\tau
,t+\tau ^{\prime }}\tilde{\Phi}_{t+\tau ^{\prime },t}^{st}[\rho _{y}])%
\mathrm{Tr}_{s}(E_{y}\Phi _{t,t^{\prime }}^{st}\Lambda _{t^{\prime },0}[%
\tilde{\rho}_{x}])]},  \label{3JointClasica}
\end{equation}%
\end{widetext}%
where here%
\begin{equation}
\tilde{\Phi}_{t_{b},t_{a}}^{st}=\mathcal{\mathcal{L}}_{st}(t_{b})\mathcal{G}%
_{t_{b},t_{a}}^{st},\ \ \ \ \ \ \ \ \Phi _{t_{b},t_{a}}^{st}\equiv \mathcal{G%
}_{t_{b},t_{a}}^{st}\mathcal{QL}_{st}(t_{a}),
\end{equation}%
and correspondingly%
\begin{equation}
\mathcal{G}_{t,t^{\prime }}^{st}=\left\lceil \exp \int_{t^{\prime
}}^{t}dt^{\prime }\mathcal{QL}_{st}(t^{\prime })\right\rceil .
\end{equation}%
The propagator $\Lambda _{t_{b},t_{a}}$ is defined by Eq. (\ref{PropaClasico}%
). The CPF correlation, using the similitude of Eqs. (\ref{P3Noise}) and (%
\ref{CPFNoise}), can be written from Eq. (\ref{3JointClasica}) as%
\begin{equation}
C_{pf}(t,\tau )|_{y}=\sum_{z,x}\frac{O_{z}O_{x}}{P^{2}(y)}%
\int_{0}^{t}dt^{\prime }\int_{0}^{\tau }d\tau ^{\prime }\overline{\Xi (z,x|y)%
}[\tilde{\rho}_{yx}],
\end{equation}%
where%
\begin{equation}
\overline{\Xi (z,x|y)}[\bullet ]\equiv \overline{\mathrm{Tr}_{s}(E_{z}\tilde{%
\Upsilon}_{t+\tau ^{\prime },t}^{st}[\rho _{y}])\mathrm{Tr}%
_{s}(E_{y}\Upsilon _{t,t^{\prime }}^{st}[\bullet ])}.  \label{SUMA}
\end{equation}%
Similarly, we defined%
\begin{equation}
\tilde{\Upsilon}_{t+\tau ^{\prime },t}^{st}[\bullet ]\equiv \Lambda _{t+\tau
,t+\tau ^{\prime }}\tilde{\Phi}_{t+\tau ^{\prime },t}^{st}[\bullet ],
\end{equation}%
and the stochastic superoperator%
\begin{equation}
\Upsilon _{t,t^{\prime }}^{st}[\bullet ]\equiv \Phi _{t,t^{\prime
}}^{st}[\Lambda _{t^{\prime },0}[\bullet ]].
\end{equation}%
Furthermore, $\rho _{yx}$ is the matrix defined by Eq. (\ref{RhoYX}). From
Eq. (\ref{SUMA}) the perturbation theory follows straightforwardly.

\section{Examples and applications}

In this section we apply the perturbation theory for different dynamics of
interest such as dephasing induced by a Gaussian non-white noise and
dissipation induced by a non-Markovian bosonic thermal bath.

\subsection{Non-Markovian dephasing}

We consider a two-level system driven by a dephasing stochastic Hamiltonian.
The stochastic system state $\rho _{t}^{st}$ evolves as%
\begin{equation}
\frac{d}{dt}\rho _{t}^{st}=-i\xi (t)[\sigma _{\hat{z}},\rho _{t}^{st}],
\label{Dephasing}
\end{equation}%
where $\sigma _{\hat{z}}$ is the $\hat{z}$-Pauli matrix (eigenvalues $|\pm
\rangle $) and $\xi (t)$ is a (real) stationary color Gaussian noise with
vanishing average $\overline{\xi (t)}=0$ and stationary correlation $\chi
(t-t^{\prime })=\overline{\xi (t)\xi (t^{\prime })}=(\gamma /2\tau _{c})\exp
[-|t-t^{\prime }|/\tau _{c}].$ We consider that the system begins in its
upper state, $\rho _{0}=|+\rangle \langle +|.$\ Furthermore, the three
measurements are performed in the $\hat{x}$-direction in the Bloch sphere, $%
\{\Omega _{x}\}=\{\Omega _{y}\}=\{\Omega _{z}\}=|\hat{x}_{\pm }\rangle
\langle \hat{x}_{\pm }|,$ where $|\hat{x}_{\pm }\rangle =(|+\rangle \pm
|-\rangle )/\sqrt{2}$ are the eigenvalues of $\sigma _{\hat{x}},$ the $\hat{x%
}$-Pauli matrix. Thus, $x=\pm 1,$ $y=\pm 1,$ and $z=\pm 1.$ Under the
previous conditions, both the joint probabilities and CPF correlation can be
obtained in an exact analytical way. Explicit expressions can be found in
Ref. \cite{budiniCPF} $[(\gamma /2\tau _{c})\leftrightarrow g^{2}].$
Similarly, for this model it is possible to obtain explicit recursive
relations and expressions for the successive series terms [Eq.~(\ref%
{SerieSigma})], which are of order $(\gamma \tau _{c})^{n}.$ Due to the
symmetry of the problem, the first order contribution vanishes.

In Fig.~1 we plot the CPF correlation $C_{pf}(t,\tau )|_{y}$ at equal
measurement time intervals $t=\tau$, for different noise correlation times.
Both the exact expression and the perturbation theory estimation are shown.
The unperturbed system propagator $\Lambda _{t,t^{\prime }}$ [Eq. (\ref%
{PropaClasico})] was taken as the exact one. Similarly to the exact
expression, the CPF correlation obtained by adding successive series terms
is independent of the conditional $y=\pm 1.$ We found that for smaller noise
correlation times the convergence to the exact expression is increased,
which shows the consistence of the perturbation theory. Furthermore, we
checked that, to the same order, all joint probabilities $P(z,y,x)$ [Eq. (%
\ref{3JointClasica})] are definite positive. This feature also supports the
developed formalism. 
\begin{figure}[tb]
\centering
\includegraphics[width=1\linewidth]{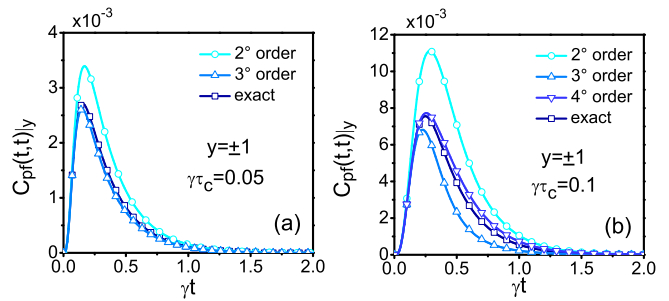}
\caption{CPF correlation $C_{pf}(t,t)|_{y}$ $(y=\pm 1)$ obtained by adding
successive contributions in the perturbation theory for the dephasing
dynamics (\protect\ref{Dephasing}). The noise correlation is $\protect\chi %
(t)=(\protect\gamma /2\protect\tau _{c})\exp [-|t|/\protect\tau _{c}].$ The
three measurements are performed in the $\hat{x}$-Bloch direction, while the
system begins in its upper state. The parameters in (a) and (b) respectively
are $\protect\gamma \protect\tau _{c}=0.05\ $and $0.1.$}
\label{FIG_1}
\end{figure}

\subsection{Non-Markovian bosonic bath}

The decay of a two-level system in a bosonic environment is described by the
total Hamiltonian \cite{breuerbook}%
\begin{equation}
H_{\mathrm{tot}}=\frac{\omega _{0}}{2}\sigma _{z}+\sum_{k}\omega
_{k}b_{k}^{\dag }b_{k}+\sum_{k}(g_{k}\sigma _{+}b_{k}+g_{k}^{\ast }\sigma
_{-}b_{k}^{\dag }),
\end{equation}%
where $[b_{k},b_{k}^{\dag }]=1$ are the creation-annihilation bosonic
operators and $\sigma _{+}=|+\rangle \langle -|,$ $\sigma _{-}=|-\rangle
\langle +|$ are the raising and lowering operators of the system. Memory
effects in this dynamics can also be analyzed in an operational approach to
quantum non-Markovianity.

We consider two different measurement schemes. In the first one, the three
measurements are performed in $\hat{z}-$Bloch direction ($\hat{z}$-$\hat{z}$-%
$\hat{z}$ scheme) while in the second one, the first and last measurements
are performed in the $\hat{x}-$Bloch direction, while the intermediate one
in the $\hat{z}$-direction ($\hat{x}$-$\hat{z}$-$\hat{x}$ scheme). These
observables are defined in the representation interaction with respect to
the system and bath free evolutions. In this frame, the total Hamiltonian
reads 
\begin{equation}
H_{\mathrm{tot}}=\sigma _{+}B(t)+\sigma _{-}B^{\dagger }(t),
\label{Dissipation}
\end{equation}%
where $B(t)=\sum_{k}g_{k}b_{k}\exp [+i(\omega _{0}-\omega _{k})t].$
Furthermore, the initial bipartite state is taken as%
\begin{equation}
\rho _{0}^{se}=|\psi _{0}\rangle \langle \psi _{0}|\otimes \sigma _{e},\ \ \
\ \ \ \ |\psi _{0}\rangle =(a|+\rangle +b|-\rangle ),
\end{equation}%
with normalized coefficients $a$ and $b.$ The initial bath state $\sigma
_{e} $ is taken as a thermal one. For both measurement schemes, the
perturbation theory enables us to study the dependence of memory effects
with temperature. Given the bosonic property of the bath, its complete set
of (operator) correlations can be written in terms of only two ones,%
\begin{equation}
\chi _{\downarrow }(t)\equiv \mathrm{Tr}_{e}[B(t)B^{\dagger }\sigma _{e}],\
\ \ \ \ \ \chi _{\uparrow }(t)\equiv \mathrm{Tr}_{e}[B^{\dagger }(t)B\sigma
_{e}].  \label{BathCorre}
\end{equation}%
%
%
%
%
%
%
%
%
%
%
%
%
%
%
%
%
%
%
%
\begin{figure}[tb]
\centering
\includegraphics[width=1\linewidth]{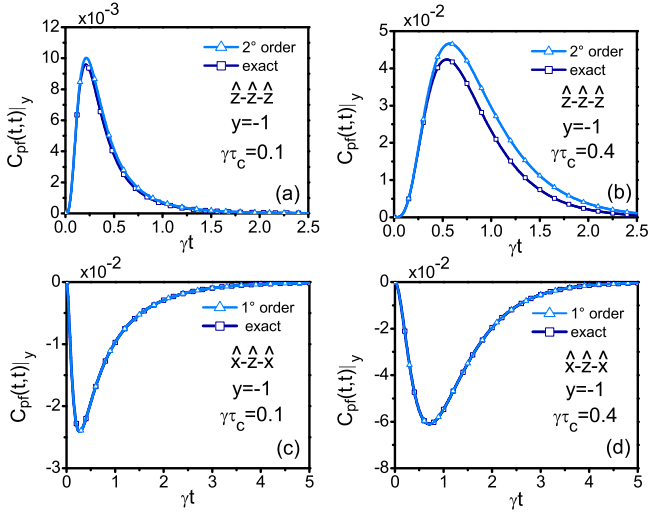}
\caption{CPF correlation $C_{pf}(t,t)|_{y}$ $(y=-1)$ obtained in a
perturbative way for the dissipative dynamics (\protect\ref{Dissipation})
with environment at zero temperature. Its correlations are $\protect\chi %
_{\downarrow }(t)=(\protect\gamma /2\protect\tau _{c})\exp [-|t|/\protect%
\tau _{c}]$ and $\protect\chi _{\uparrow }(t)=0.$ (a) and (b) correspond to
the $\hat{z}$-$\hat{z}$-$\hat{z}$ measurement scheme while (c) and (d) to
the $\hat{x}$-$\hat{z}$-$\hat{x}$ scheme. The bath correlation time $\protect%
\tau _{c}$ is indicated in each plot. In all cases, the initial system state
is $|\protect\psi _{0}\rangle =(\protect\sqrt{p}|+\rangle +\protect\sqrt{1-p}%
|-\rangle )$ with $p=0.8.$}
\label{FIG_2}
\end{figure}

\textit{Zero temperature}: For different physical arrangements, the
environment temperature can be (effectively) taken as null. Thus, $\sigma
_{e}=|0\rangle \langle 0|,$ where $|0\rangle \equiv \prod_{k}|0\rangle _{k}.$
Each state $|0\rangle _{k}$ corresponds to the vacuum state of each bosonic
mode. As is well known \cite{breuerbook}, in this case the full
system-environment dynamic admits a simple analytical solution, given also
an exact expression for the unperturbed system propagator $\Lambda
_{t,t^{\prime }}$\ [Eq.~(\ref{PropaCuantico})]. Furthermore, the CPF
correlation and joint probabilities can also be calculated in an exact way.
In fact, the open system dynamics and the CPF correlation have been
implemented and measured in a photonic setup \cite{budiniBrasil}.

We consider a Lorentzian spectral bath density. Thus, the environment
correlations read $\chi _{\downarrow }(t)=(\gamma /2\tau _{c})\exp
[-|t|/\tau _{c}],$ while the zero temperature condition leads to $\chi
_{\uparrow }(t)=0.$ In Fig.~2 we plot the CPF correlation $C_{pf}(t,t)|_{y}$
at equal time intervals for both measurement schemes and the conditional $%
y=-1.$ In the $\hat{z}$-$\hat{z}$-$\hat{z}$ scheme [(a) and (b)], the first
order contribution vanishes. Similarly to the previous case, (at second
order) a decrease in the bath correlation time leads to a higher convergence
with the exact analytical result \cite{budiniBrasil}. On the other hand, in
the $\hat{x}$-$\hat{z}$-$\hat{x}$ scheme, the first order contribution
coincides with the exact solution. Thus, while higher order contributions do
not vanish, their addition cancel out. These results also support the
consistence of the perturbation theory. In addition, we found that to the
same order, all joint probabilities $P(z,y,x)$ [Eq.~(\ref{3JointQuantum})]
are definite positive.

For the conditional $y=+1,$ the exact calculation of the CPF correlation
leads to \cite{budiniBrasil}%
\begin{equation}
C_{pf}(t,\tau )|_{y=+1}\underset{\hat{z}\hat{z}\hat{z}}{=}0,\ \ \ \ \ \ \ \
C_{pf}(t,\tau )|_{y=+1}\underset{\hat{x}\hat{z}\hat{x}}{=}0.  \label{y=+1}
\end{equation}%
In this case, the joint probabilities $P(z,y,x)$ can be written as $P(z,1,x)%
\underset{\hat{z}\hat{z}\hat{z}}{=}P(z|1)P(1|x)P(x)$and $P(z,1,x)\underset{%
\hat{x}\hat{z}\hat{x}}{=}P(z)P(1)P(x)$ \cite{probability}. In fact, these
expressions correspond to the first contribution in Eq. (\ref{3JointQuantum}%
), while the integral contribution vanishes. Thus, in this restricted case $%
(y=+1),$ the Markov property is fulfilled leading to a vanishing CPF
correlation and, consequently, the perturbation theory loses its meaning.%
\begin{figure}[tb]
\centering
\includegraphics[width=1\linewidth]{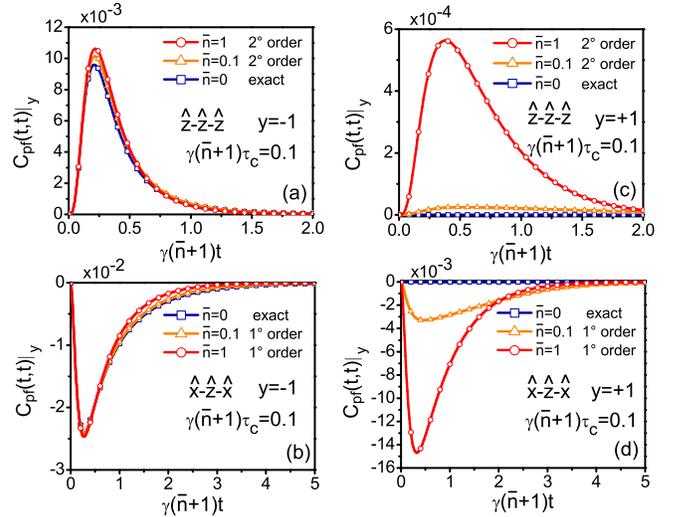}
\caption{CPF correlation $C_{pf}(t,t)|_{y}$ obtained in a perturbative way
for the dissipative dynamics (\protect\ref{Dissipation}) with an environment
at finite temperature. Its correlations are $\protect\chi _{\downarrow }(t)=(%
\bar{n}+1)(\protect\gamma /2\protect\tau _{c})\exp [-|t|/\protect\tau _{c}]$
and $\protect\chi _{\uparrow }(t)=\bar{n}(\protect\gamma /2\protect\tau %
_{c})\exp [-|t|/\protect\tau _{c}].$ (a) and (b) correspond to $y=-1$ for
the $\hat{z}$-$\hat{z}$-$\hat{z}$ and $\hat{x}$-$\hat{z}$-$\hat{x}$
measurement schemes, respectively. In (c) and (d) $y=+1.$ The bath
correlation parameters are indicated in each plot. In all cases, the initial
system state $|\protect\psi _{0}\rangle $ is the same as in Fig.~2.}
\label{FIG_3}
\end{figure}

\textit{Finite temperature}: For finite temperature, a simple expression for
the unperturbed system propagator is not available. In addition, neither the
CPF correlation nor the joint probabilities can be obtained in an exact
analytical way. Nevertheless, this case can be dealt with the developed
perturbation theory.

At finite temperature, both bath correlations [Eq.~(\ref{BathCorre})] must
be considered. As a model, we take $\chi _{\downarrow }(t)=(\bar{n}%
+1)(\gamma /2\tau _{c})\exp [-|t|/\tau _{c}],$ and $\chi _{\uparrow }(t)=%
\bar{n}(\gamma /2\tau _{c})\exp [-|t|/\tau _{c}],$ where $\bar{n}$\ is the
average number of bosonic bath excitations at the natural frequency of the
system. When $\bar{n}=0$\ the previous Lorentzian case at null temperature
is recovered. This correlation model arises when the dependence on frequency
of the number of thermal bath excitations is almost a flat function around
the natural system frequency \cite{mandel}. In this approximation,
temperature increases the \textquotedblleft intensity\textquotedblright\ of
the environment fluctuations, while their correlation time is independent of
it. Consistently, the unperturbed density matrix propagator is taken as the
(exact) zero temperature propagator \cite{breuerbook}, with an extra similar
contribution that takes into account (thermally induced) transitions from
the lower to the upper system state.

By using the previous assumptions, in Fig.~3 we plot the CPF correlation
obtained from the perturbation theory [Eqs. (\ref{Final}) and (\ref%
{SerieSigma})]. For both measurement schemes ($\hat{z}$-$\hat{z}$-$\hat{z}$
and $\hat{x}$-$\hat{z}$-$\hat{x}$), and for the conditional $y=-1$
[Figs.~3(a) and (b)], the memory effects (amplitude of the CPF correlation)
weakly depends on temperature. Small departures with respect to the
vanishing temperature case [Fig.~2] are observed. Due to the normalization
of the time axis $[\gamma (\bar{n}+1)t],$ a natural change of time scale
(shrinking due to the increasing of the effective system decay rates) is not
observed.

On the other\ hand, for the conditional $y=+1$ [Figs.~3(c) and (d)] a strong
dependence on temperature is observed in both measurement schemes. In fact,
in this situation, in the limit $\bar{n}\rightarrow 0,$ the CPF correlation
vanishes, Eq.~(\ref{y=+1}). By increasing temperature the maximal amplitude
of the CPF correlation also increases.

The previous unusual effect, that is, an increasing of the memory effects
with temperature, does not rely on the specific environmental properties
such as the proposed correlation model. It relies on the symmetries of the
problem, which are defined by the system-environment interaction and the
quantum measurement\ processes. For the conditional $y=+1,$ an increasing of
the bath temperature leads to an extra dissipation channel that breaks the
conditional statistical independence of the first and last (past and future)
measurement outcomes. From the point of view of joint probabilities,
temperature leads to extra contributions that break the Markovian property.
In fact, for $y=+1$ we found that the first integral series contributions in
Eq.~(\ref{3JointQuantum}) are proportional to $\bar{n},$ while in the
previous case $(y=-1)$ are proportional to $(\bar{n}+1).$ We checked that by
increasing temperature, the memory effects saturates. In addition, a
vanishing of memory effects with temperature can be introduced through a
temperature dependent bath correlation time.

\subsection{Generalizations}

The present approach is generalizable to different cases of interest. First,
the formalism can be extended by considering arbitrary initial conditions, $%
\rho _{0}^{se}\neq \rho _{0}\otimes \sigma _{e}.$ For example the bath state
can be an arbitrary one, different from the reference state in the
projectors definition [Eq.~(\ref{QProjectores})]. In addition system and
environment may be correlated at the initial time. These situations lead to
an extra term in the perturbation theory [see Eq.~(\ref{SolIrrelevant}) in
the Appendix]. On the other hand, higher statistical objects can also be
worked out with similar techniques. In fact, one may consider higher joint
probabilities $P(x_{m},\cdots x_{1})$ involving $m$-measurement processes 
\cite{modi}. In this case, $(m-1)$ system propagators are involved in the
convolution term, while the structure of series terms assumes a similar
form. The same result applies to higher CPF correlations \cite{budiniCPF}.

\section{Summary and Conclusions}

Using projector techniques, we have developed a perturbation theory for
describing memory effects in operational approaches to quantum
non-Markovianity, where the system dynamics is explicitly observed at
different times. The formalism leads to exact expressions of both joint
probabilities and correlations, which are written in terms of the
unperturbed system density matrix propagator. We worked out the minimal case
of three measurement processes. Memory contributions are defined by a
convolution integral involving two system propagators, where the successive
series terms are weighted by higher order bath correlations. In a bosonic or
Gaussian case they can be reduced to two-point correlations. This result
clarifies which structure determines memory effects in operational
approaches to quantum non-Markovianity.

As examples, we applied the theory to different open system dynamics that
admit an exact treatment, such as dephasing induced by stochastic
Hamiltonians and decay of a two-level system in a bosonic reservoir at zero
temperature. The consistence between the perturbation theory and exact
solutions shows and guarantees the validity of the proposed approach. We
also studied non-Markovian effects that emerge when considering thermal
baths. Unusual memory effects arise due to the interplay between the
measurement process, the bipartite dynamics, and the environment
temperature. We found that, depending on the chosen measurement processes
and conditionals, memory effects may grow with the environment temperature.
This feature can be understood from a special interplay between the previous
ingredients, where an extra dissipative channel induced by the environment
temperature facilitates the developing of memory effects.

To conclude, our theory provides a solid basis for analyzing memory effects
in operational approaches to quantum non-Markovianity. Application to others
physical arrangements and statistical objects can be tackled by using the
developed formalism.

\section*{Acknowledgments}

M.B. thanks support from Comisi\'{o}n Nacional de Energ\'{\i}a At\'{o}mica
(CNEA), Argentina. A.A.B. thanks support from Consejo Nacional de
Investigaciones Cient\'{\i}ficas y T\'{e}cnicas (CONICET), Argentina.

\appendix*

\section{Formal solutions from projector techniques}

As usual \cite{projectors}, a bipartite system-environment evolution, $%
(d/dt)\rho _{t}^{se}=\mathcal{L}_{se}(t)\rho _{t}^{se},$ can be split as 
\begin{subequations}
\label{Projectores}
\begin{eqnarray}
\frac{d}{dt}\mathcal{P}\rho _{t}^{se} &=&\mathcal{PL}_{se}(t)\mathcal{P}\rho
_{t}^{se}+\mathcal{PL}_{se}(t)\mathcal{Q}\rho _{t}^{se}, \\
\frac{d}{dt}\mathcal{Q}\rho _{t}^{se} &=&\mathcal{QL}_{se}(t)\mathcal{P}\rho
_{t}^{se}+\mathcal{QL}_{se}(t)\mathcal{Q}\rho _{t}^{se},
\end{eqnarray}%
where the projectors\ $\mathcal{P}$\ and $\mathcal{Q}$\ are given by Eq. (%
\ref{QProjectores}). The irrelevant part $\mathcal{Q}\rho _{t}^{se}$\ can be
integrated as 
\end{subequations}
\begin{equation}
\mathcal{Q}\rho _{t}^{se}=\mathcal{G}_{t,t_{0}}\mathcal{Q}\rho
_{t_{0}}^{se}+\int_{t_{0}}^{t}dt^{\prime }\mathcal{G}_{t,t^{\prime }}%
\mathcal{QL}_{se}(t^{\prime })\mathcal{P}\rho _{t^{\prime }}^{se},
\label{SolIrrelevant}
\end{equation}%
where the corresponding propagator is%
\begin{equation}
\mathcal{G}_{t,t^{\prime }}=\left\lceil \exp \int_{t^{\prime
}}^{t}dt^{\prime }\mathcal{QL}_{se}(t^{\prime })\right\rceil .
\end{equation}%
By introducing the solution Eq. (\ref{SolIrrelevant}) into Eq. (\ref%
{Projectores}) the evolution of the relevant part $\mathcal{P}\rho _{t}^{se}$
can be written as%
\begin{equation}
\frac{d}{dt}\mathcal{P}\rho _{t}^{se}=\int_{t_{0}}^{t}dt^{\prime }\mathbb{K}%
(t,t^{\prime })\mathcal{P}\rho _{t^{\prime }}^{se}+\mathbb{I}_{t,t_{0}},
\label{EvolutionRelevant}
\end{equation}%
where the exact memory kernel is%
\begin{equation}
\mathbb{K}(t,t^{\prime })=\delta (t-t^{\prime })\mathcal{PL}_{se}(t)+%
\mathcal{PL}_{se}(t)\mathcal{G}_{t,t^{\prime }}\mathcal{QL}_{se}(t^{\prime
}),
\end{equation}%
while the inhomogeneous term is%
\begin{equation}
\mathbb{I}_{t,t_{0}}=\mathcal{PL}_{se}(t)\mathcal{G}_{t,t_{0}}\mathcal{Q}%
\rho _{t_{0}}^{se}.  \label{Inhomogeneo}
\end{equation}%
The evolution (\ref{EvolutionRelevant}) can be solved in a formal way by
noticing that the solution of the homogeneous part can be written as $%
\mathcal{PE}_{t,t_{0}}\mathcal{P}\rho _{t_{0}}^{se}.$ Thus, the full
solution is given by%
\begin{equation}
\mathcal{P}\rho _{t}^{se}=\mathcal{PE}_{t,t_{0}}\mathcal{P}\rho
_{t_{0}}^{se}+\int_{t_{0}}^{t}dt^{\prime }\mathcal{PE}_{t,t^{\prime }}%
\mathcal{PL}_{se}(t^{\prime })\mathcal{G}_{t^{\prime },t_{0}}\mathcal{Q}\rho
_{t_{0}}^{se},  \label{SolRelevant}
\end{equation}%
where we have used the explicit expression for the inhomogeneous
contribution, Eq.~(\ref{Inhomogeneo}). The general solution Eqs. (\ref%
{SolIrrelevant}) and (\ref{SolRelevant}) support Eqs. (\ref{IrrelevantPart})
and (\ref{RelevantPart}) respectively.


\begin{thebibliography}{99}
\bibitem{probability} These results can straightforwardly be derived by
using that $P(z,y,x)=P(z,x|y)P(y),$ where $P(y)=\sum_{x}P(y|x)P(x),$ and
using the exact probability expressions of Ref.~\cite{budiniBrasil}.
\end{thebibliography}
\end{document}